Two-level model [1] is widely used to describe thermodynamic properties of amorphous materials. Accordingly two states for each atom of a system are considered: a ground state with zero energy, and an excited state with energy $E$, the value of which depends on the distribution of surrounding atoms. The fraction of excited atoms in a state with energy $E$ is determined by the Boltzmann equation,

$$f = (1 - f)\exp(-E/T), \qquad (1)$$

where $E$ is in the units of the Boltzmann constant, $k$.

Let us denote the fraction of the atoms, having energy between $E$ and $E + dE$ as $W(E)dE$. Then the total energy of a system is given by the following integral from zero to infinity:

$$\langle E \rangle = \int EW(E)dE/(1 + \exp E/T) = T^2 \int xW(xT)dx/(1 + \exp x). \qquad (2)$$

At low enough temperature only some small vicinity near $T = 0$ contributes to the integral in the right-hand part of Eq. 2. When the density of states is analytical function in the vicinity of $E = 0$, it is possible to approximate $\langle E \rangle$ as follows [1, 2]:

$$\langle E \rangle = T^2 W(0) \int x dx/(1 + \exp x) \equiv T^2/2T_0, \qquad (3)$$

with $T_0 = 6/\pi^2 W(0)$.

A well know expression for the heat capacity (per atom, and divided by the Boltzmann constant) follows immediately from Eq. (3):

$$c = T/T_0. \qquad (4)$$

In a general case an excited state of the atom is separated by some barrier from the ground state, however. Let $E_1$ denotes the height of the barrier. For transition from the excited state to the ground one, it is necessary to overcome barrier of the height $E_1$. For transition from ground state to the excited one, a barrier of the height $E + E_1$ should be overcome. This process takes some time. The barrier could be overcome by the mechanism of thermal fluctuations, and by tunneling at low temperatures. From this follows that the measured heat capacity depends on the rate of the temperature variation and on the initial state of a system. To investigate this dependence the kinetic equation should be written down. The fraction of the atoms, leaving ground state for the excited one in unit time is $\nu(1 - f)[a(\exp -E/T) + \exp(-E - E_1)/T]$ (here $\nu$ is some factor, the value of which is about the frequency of the oscillation of the atom, $a$ characterizes the permeability of the barrier). We assume here that $a$ is independent on $E$. The fraction of the atoms, leaving excited state for the ground one in unit time is $\nu f[a + (\exp -E_1/T)]$. The kinetic equation is as follows:

$$df/dt + (f - f_0)/\tau = 0,$$
$$f_0 = (1 + \exp E/T)^{-1}, \quad 1/\tau(E) = \nu(a + \exp -E_1/T)(1 + \exp -E/T). \qquad (5)$$

Enthalpy per atom is $h = fE$. Multiplication of the obtained kinetic equation by $E\tau(E)$ with subsequent averaging on $E$ with the weight $W(E)$ yields:

$$T_t \langle \tau(E)dh/dT \rangle + \langle h \rangle = \langle h_0 \rangle,$$

Where $T_t = dT/dt$, and $h_0$ corresponds to the equilibrium state.

It follows from Eq. (5) that variation of $E$ from zero to infinity causes two times changes in $\tau(E)$ only. Therefore it is convenient to use the theorem of average and write instead of $\tau(E)$ some average value of it, which we will further on denote as $\tau$. Thus for low temperatures we have

$$T_t \tau (d\langle h \rangle/dT) + \langle h \rangle = T^2/2T_0. \tag{6}$$

When $T$ is much smaller than $E_1/\ln(1/a)$, $1/\tau = Cva$ ($1 \leq C \leq 2$) and does not depend on $T$. Eq. (6) determines variation of $\langle h \rangle$ from its initial value.

Let us consider example when temperature changes from 0 to $T$ with some constant rate $T_t > 0$. Let the initial enthalpy was zero. In this case

$$\langle h \rangle = (T^2/2T_0) - (T_t\tau/T_0)T + (T_t\tau)^2[1 - \exp(-T/T_t\tau)]/T_0. \tag{7}$$

Measured dynamic heat capacity is given by the following expression

$$c = (T/T_0) - (T_t\tau/T_0)[1 - \exp(-T/T_t\tau)]. \tag{8}$$

Comparing Eqs. (8) and (4) shows that the dynamic heat capacity is smaller than the static one. At low heating rates, when $T_t\tau$ is essentially smaller than $T$, both heat capacities are equal to each other. When $T_t\tau$ is essentially larger than $T$, the dynamic heat capacity is much smaller than the static one:

$$c = T^2/2T_0T_t\tau. \tag{9}$$

Eq. (9) allows determining the relaxation time $\tau$. Then $\tau$ could be used to calculate the static heat capacity $T/T_0$. For this we have to take into account data on measured heat capacity, described by Eq. (8).

As the relaxation time $\tau$ could be extremely large for amorphous materials at low temperature, peculiarities, considered in present communication, look very essential.

1. *Anderson P. W., Halperin B. I., Varma C. M.*. Anomalous low-temperature thermal properties of glasses and spin-glasses.–Phil. Mag., 1972, **25**, N 1, p. 1-9.
2. *Phillips W. A*. Tunneling states in amorphous solids.-J. Low Temp. Phys., 1972, **7**, N 3/4, p. 351-360.